\documentclass[sigconf]{acmart}

\setcopyright{none}
\renewcommand\footnotetextcopyrightpermission[1]{} 
\usepackage{multirow}
\usepackage{graphicx} 
\usepackage{float}
\usepackage{subfigure}
\usepackage[normalem]{ulem}
\useunder{\uline}{\ul}{}
\usepackage{amsmath}
\usepackage{array}

\AtBeginDocument{%
  \providecommand\BibTeX{{%
    \normalfont B\kern-0.5em{\scshape i\kern-0.25em b}\kern-0.8em\TeX}}}

\begin{document}

\title{Efficient Multi-task Prompt Tuning for Recommendation}

\author{Ting Bai}
\affiliation{%
  \institution{Beijing University of Posts and Telecommunications}
  \country{Beijing, China}
}
\email{baiting@bupt.edu.cn}

\author{Le Huang}
\affiliation{%
  \institution{Beijing University of Posts and Telecommunications}
  \country{Beijing, China}
}
\email{lehuang@bupt.edu.cn}

\author{Yue Yu}
\affiliation{%
  \institution{Beijing University of Posts and Telecommunications}
  \country{Beijing, China}
}
\email{loadingyy@bupt.edu.cn}

\author{Cheng Yang}
\affiliation{%
  \institution{Beijing University of Posts and Telecommunications}
  \country{Beijing, China}
}
\email{yangcheng@bupt.edu.cn}

\author{Cheng Hou}
\affiliation{%
  \institution{Tencent AI Lab}
  \country{Beijing, China}
}
\email{chenghou@tencent.com}

\author{Zhe Zhao}
\affiliation{%
  \institution{Tencent AI Lab}
  \country{Beijing, China}
}
\email{nlpzhezhao@tencent.com}

\author{Chuan Shi}
\affiliation{%
  \institution{Beijing University of Posts and Telecommunications}
  \country{Beijing, China}
}
\email{shichuan@bupt.edu.cn}

\begin{abstract}
With the expansion of business scenarios, real recommender systems are facing challenges in dealing with the constantly emerging new tasks in multi-task learning frameworks.
In this paper, we attempt to improve the generalization ability of multi-task recommendations when dealing with new tasks. We find that joint training will enhance the performance of the new task but always negatively impact existing tasks in most multi-task learning methods. Besides, such a re-training mechanism with new tasks increases the training costs, limiting the generalization ability of multi-task recommendation models. Based on this consideration, we aim to design a suitable sharing mechanism among different tasks while maintaining joint optimization efficiency in new task learning. A novel two-stage prompt-tuning MTL framework (MPT-Rec) is proposed to address task irrelevance and training efficiency problems in multi-task recommender systems. Specifically, we disentangle the task-specific and task-sharing information in the multi-task pre-training stage, then use task-aware prompts to transfer knowledge from other tasks to the new task effectively. By freezing parameters in the pre-training tasks, MPT-Rec solves the negative impacts that may be brought by the new task and greatly reduces the training costs. Extensive experiments on three real-world datasets show the effectiveness of our proposed multi-task learning framework. MPT-Rec achieves the best performance compared to the SOTA multi-task learning method. Besides, it maintains comparable model performance but vastly improves the training efficiency (i.e., with up to 10\% parameters in the full training way) in the new task learning.
\end{abstract}


\maketitle

\section{Introduction}

\begin{figure}[h]
  \centering
  \includegraphics[width=0.8\linewidth]{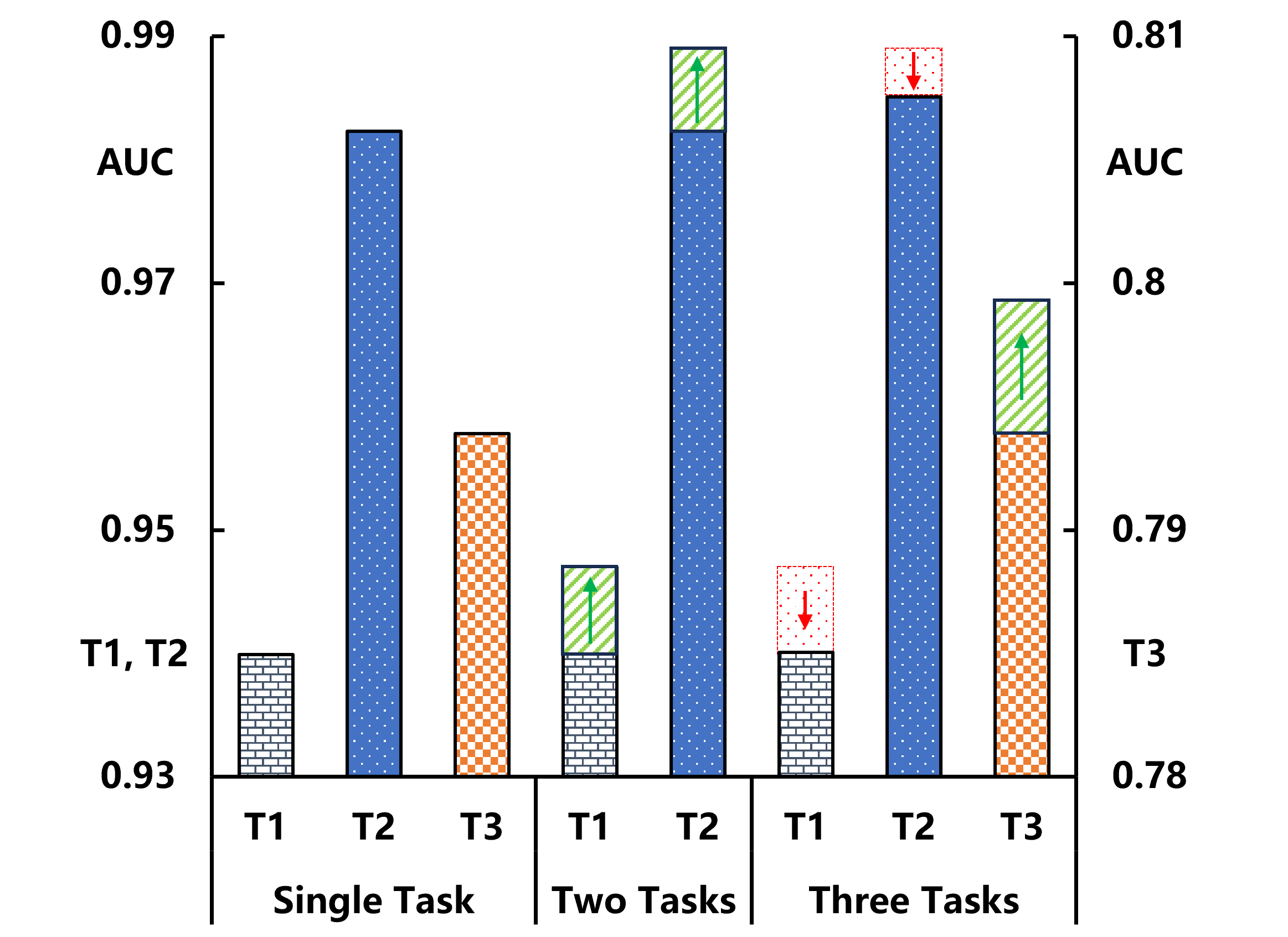}
  \caption{The experimental results of multi-task learning method MMOE on Census-income dataset. Task T1, T2 and T3 are the predictions of the "income", "marital status", and "sex" labels. We can see that compared with the AUC performance on the single task, MTL on two tasks promotes each other. Learning with the new task T3, MTL improves the performance on new task T3, but damages T1 and T2 which had been optimized in the two-tasks stage. }
\label{fig:Introduction}
\end{figure}

Multi-task learning (MTL) refers to optimizing different tasks together to make full utilization of the information contained in other tasks.
Due to its broad application in recommender systems,
for example, the Click Through Rate (CTR) and Click Conversion Rate (CVR) are usually jointly optimized,
multi-task learning has become an active research topic in recent recommendation studies~\cite{collobert2008unified,luong2015multi,zamir2018taskonomy,lu2018like, thung2018brief, zhao2019recommending, crawshaw2020multi}. 
Different from transfer learning~\cite{pan2009survey, weiss2016survey, zhuang2020comprehensive}, which uses the associated tasks to provide extra information to the main task, multi-task learning aims to optimize all tasks at the same time. 
To achieve better performance on all different tasks, existing studies on multi-task learning~\cite{ma2018modeling, tang2020progressive, bai2022contrastive, su2023stem} mainly focus on designing effective knowledge-sharing mechanisms to avoid the negative transfer problem, i.e., transferring unrelated information from other tasks. 
Most of them ignore their generalization ability in multi-task learning to deal with new tasks.

We find that not all new tasks will promote the overall performance in existing multi-task learning frameworks. 
As shown in Fig.~\ref{fig:Introduction}, compared with the joint training of existing tasks T1 and T2 in the typical multi-task learning method MMOE~\cite{ma2018modeling}, the joint optimization with new task T3 promotes the performance on single task T3, but it harms the efforts on the previous joint training stage (i.e., two tasks) of T1 and T2. 
To avoid such degradations of model performance and expand the utility of multi-tasking learning when dealing with new tasks,
two issues need to be urgently addressed: (1) The generalization of new tasks, especially for the irrelevant ones. If the data distribution in the new task is not consistent with existing tasks, the joint optimization strategy in MTL will give rise to the negative transfer problem, which damages model performance on existing tasks; (2) the high costs of the full training process, that is to say, all tasks ( i.e., the new task and existing tasks) will be fully re-trained with all parameters updated in MTL methods.
Such a full training process will result in low training efficiency, especially when frequent requests for new task optimization are made due to changes in business scenarios.

Inspired by the efficiency approaches in the NLP area that use prompt tuning to deal with downstream tasks based on the pre-trained language models~\cite{geng2022recommendation, deng2023unified, chu2023leveraging}, we design a Multi-task Prompt-tuning framework, termed as \textbf{MPT-Rec} to address the generalization problem in dealing the new task in recommender systems.
We aim to extract useful information from existing pre-trained tasks to promote new task learning, which enables us to avoid the negative transfer from the new task and to speed up the training process.
To achieve these goals, as shown in Fig.~\ref{fig:framework},
a two-stage learning framework is designed in MPT-Rec: multi-task pre-training stage and multi-task prompt-tuning stage. 
In the multi-task pre-training stage, we design a task-aware generative adversarial network to separate the task-sharing and task-specific information, so as to ensure the high-quality transfer of task-sharing information in the prompt learning process for new tasks. 
In the prompt-tuning stage, MPT-Rec freezes the parameters in the pre-training model and only updates the parameters of the new task with the useful knowledge transferred from the pre-trained tasks by a prompt mechanism.
By splitting multi-task learning into pre-training and prompt-tuning processes, our proposed framework MPT-Rec addresses the negative transfer and high-cost training problems in multi-task learning of new tasks.  
Our contributions are as follows:

\begin{itemize}

\item We design a novel MTL framework MPT-Rec, in which a task-aware generative adversarial network is used to separate the task-sharing and task-specific information, making it flexible to fully utilize the information from other tasks to improve the model performance in multi-task learning.

\item We also study the generalization ability of multi-task recommendations to deal with new tasks. A novel two-stage pre-training and prompt-tuning MTL framework is proposed to solve the negative transfer and high-training cost problems in the optimization process of new tasks in recommender systems.

\item 
Extensive experiments on three real-world datasets show the effectiveness of our proposed multi-task learning framework MPT-Rec. It achieves the best model performance compared to the SOTA multi-task learning method, i.e., CSRec. Besides, it vastly improves the efficiency of training in dealing with new tasks.
Compared with training our method in the full training scheme, our fine-tuning method MPT-Rec only needs up to 10\% of the parameters.

\end{itemize}

\section{related work}

\subsection{Multi-Task Learning}
In the field of recommendation systems, multi-task learning has been a popular research topic in recent years~\cite{lu2018like, luong2015multi, zamir2018taskonomy, thung2018brief, zhao2019recommending, crawshaw2020multi, wang2023multi, zhang2023advances}. It optimizes multiple tasks simultaneously to reduce training time and use the knowledge between related tasks to improve the model's performance. 
Generally, the multi-task learning methods in recommender systems can be classified into four types~\cite{bai2022contrastive}: hard sharing, soft sharing, expert sharing, and sparse sharing according to the sharing mechanism of learning parameters. 
The hard parameter sharing methods \cite{collobert2008unified, liu2019multi, caruana1997multitask, ruder2017overview} ensure the unimpeded flow of knowledge between tasks through the sharing network at the bottom, but when dealing with weakly related or unrelated tasks, it faces the negative transfer problem. 
Different from hard parameter sharing, soft parameter sharing methods~\cite{duong2015low, misra2016cross} train a separate model for each task. They achieve a sharing mechanism by adding the distance between parameters of different tasks to the joint optimization function. Although soft sharing methods perform better in dealing with weakly related tasks since they use separate parameters to realize knowledge transfer without considering task correlations, they suffer the costs of larger parameter store space and lower inference efficiency in recommender systems.
The expert sharing methods~\cite{ma2018modeling, tang2020progressive} are further proposed to solve the negative transfer problem and the seesaw phenomenon by combining the outputs of different experts with gated networks.  
Recently, some methods also consider the efficiency problem in multi-task learning. Sparse parameter sharing~\cite{sun2020learning, bai2022contrastive, upadhyay2023less} methods learn different subnets for each task, and knowledge transfer is realized through the overlapping part of subnets, so as to achieve the purpose of parameter efficiency. 
In recommender systems, CSRec~\cite{bai2022contrastive} uses contrastive learning to evaluate the influence of parameters on specific tasks and solves the problem of parameter conflict.
In our work, we propose a novel MTL framework MPT-Rec, in which a task-aware generative adversarial network is designed to separate task-sharing and task-specific information. In MPT-Rec,  the information from other tasks can be fully utilized, alleviating the negative transfer problem and improving the model performance in multi-task learning. 


\subsection{Multi-Task Generalization}
As recommendation scenarios become increasingly complicated, increasing attention has been paid to the generalization ability of multi-task learning in recent years. 
Different from transfer learning or domain adaptation~\cite{pan2009survey, weiss2016survey, zhuang2020comprehensive,wu2019understanding}, which uses the associated tasks to provide extra information to the main task, multi-task learning aims to optimize all tasks at the same time. 
To achieve better performance on the new task, some model-agnostic methods are proposed in new task adoption~\cite{finn2017model, peng2021learning,wei2020fast,luo2023mamdr}. For example, meta-learning approaches like MAML~\cite{finn2017model} are proposed to train the model’s initial parameters such that the model has maximal performance on a new task after the parameters have been updated with a small amount of data from that new task. However, in addition to learning the initializations, the key to improving the model performance in MTL is learning the effective transfer of knowledge from other tasks to the new task, which has not been well addressed in the meta-learning approaches. Besides, how to update all tasks efficiently in the new task learning process has been rarely studied in the literature on multi-task recommendations. 
As for improving the generalization ability of multi-task learning methods, it is essential to distinguish the task-specific and task-sharing information among different tasks and then transfer the helpful knowledge to the new task. From the perspective of task-specific information learning, existing MTL models can be generally classified into tower-level, gate-level and expert-level respectively~\cite{su2023stem}. For example, the Shared Bottom is a tower-level multi-task learning model, in which the parameters in the independent tower of each task are updated by the supervised task loss. MMOE is the gate-level model that learns task-specific information from the gate networks. PLE learns task-specific information also at the expert-level by the task-specific and task-sharing experts. 
Recently, an embedding-level method~\cite{su2023stem} is proposed to learn task-specific and task-sharing information. Specifically, each task has its own embeddings by simple separate embedding networks to solve the negative transfer problem. 
Our multi-task framework MPT-Rec learns task-specific information at the expert level, but more effectively by using generative advertisement neural networks.

\begin{figure*}[]
  \centering
  \includegraphics[width=0.85\linewidth]{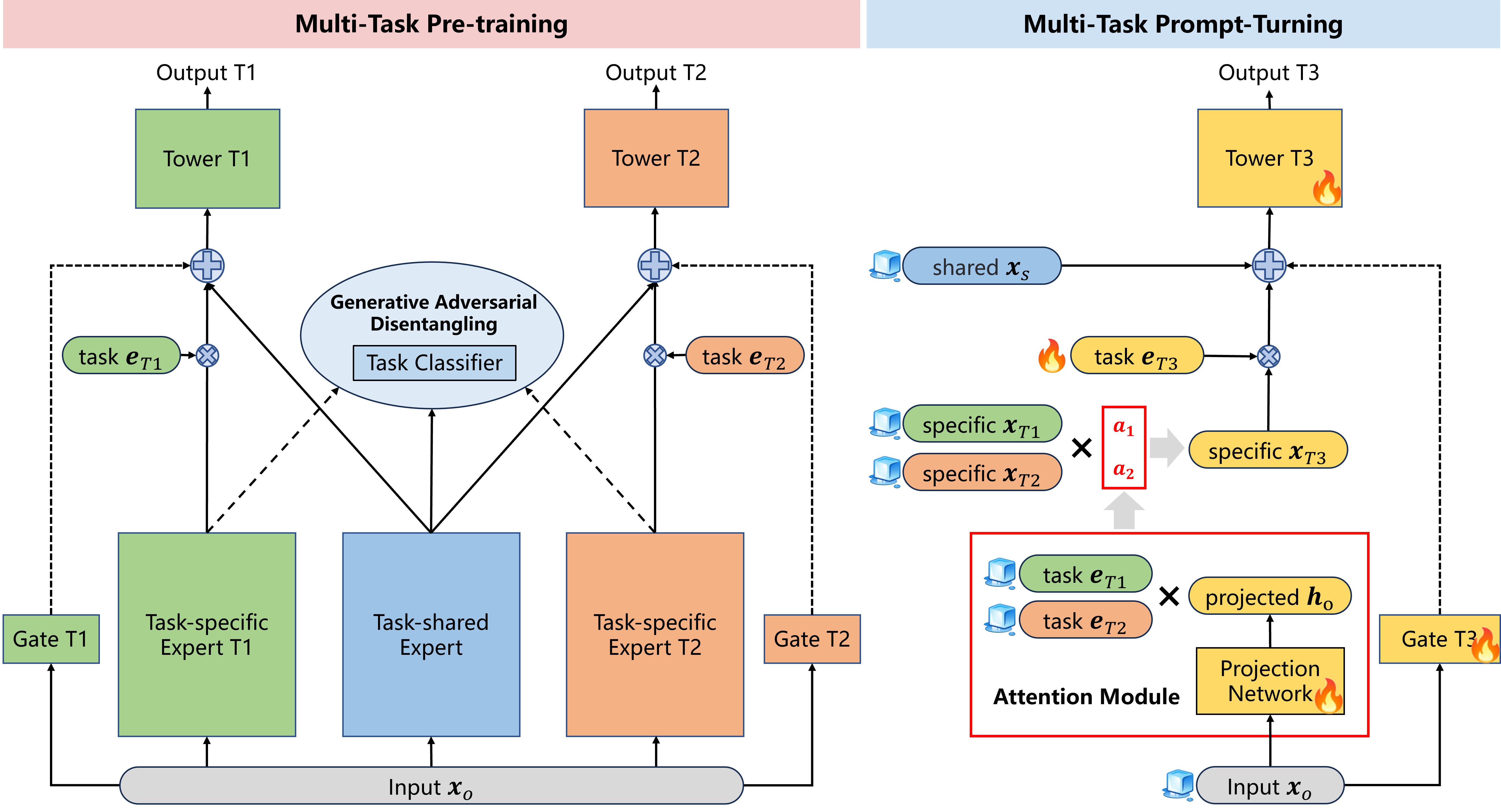}
  \caption{The overall architecture of our proposed MTL framework MPT-Rec. It consists of two components: the multi-task pre-training component and the multi-task prompt-tuning component.
  In the pre-training component, a generative adversarial network is designed to disentangle the task-specific and task-sharing information by using the task-sharing expert as a generator and the task classifier as the discriminator. In the prompt-tuning component, the parameters in the pre-training model are frozen, and useful knowledge is transferred to the new task by a prompt mechanism. Whether parameters need to be trained is indicated by labeling them as "ice" or "fire".
  }
  \label{fig:framework}
\end{figure*}

\subsection{Multi-Task Fine-tuning}
In the field of recommender systems, multi-task fine-tuning is a promising technique to make the training process more efficient~\cite{ding2023parameter,houlsby2019parameter,wang2023plate,deng2023unified,hao2024motif,zhong2024panda,zhang2024m3oe}. Its core objective is to enhance the model's generalization ability at a relatively low training cost. 
The fine-tuning operations can be implemented by transfer learning and prompt learning. 
Different from multi-task learning, which leverages knowledge among tasks to improve the overall performance of all tasks, transfer learning places greater emphasis on utilizing the similarities between source and target domains to enhance performance, specifically in the target task~\cite{zhao2020catn, thung2018brief, pan2009survey, weiss2016survey, zhuang2020comprehensive}.
Prompt learning is a method that is widely used in natural language processing (NLP) models. Its goal is to get the desired output or result by providing clear guidance or prompts. This approach is useful for many NLP tasks, including text generation, question answering, machine translation, and information retrieval \cite{liu2023hierarchical,sun2023all}.
For example, MPT \cite{wang2023multitask} learns a single transferable prompt by distilling knowledge from multiple task-specific source prompts, then learns multiplicative low-rank updates to this shared prompt to efficiently adapt it to each downstream target task.
In addition to the textual prompts, recent studies~\cite{asai2022attempt,liu2023hierarchical} also utilize embedding vectors as soft prompts. ATTEMPT \cite{asai2022attempt} combines knowledge transferred across different tasks via a mixture of soft prompts while keeping the original LM unchanged. 
However, in the field of recommendation, most of the work related to prompt learning requires a language model as an intermediary, and the prompt exists in the form of text~\cite{deng2023unified, chu2023leveraging, geng2022recommendation}. 
Different from them, our work uses task embeddings as prompt vectors, which guide the generation of task-specific representations in the new task generalization phase.  Our proposed prompt-tuning multi-task learning method has a unique advantage in terms of training efficiency.

\section{Methodology}

In this section, we briefly introduce the architecture of MPT-Rec and then explain each component in detail.

\subsection{The General Framework}
MPT-Rec consists of two components: the multi-task pre-training component and the multi-task prompt-tuning component, corresponding to the two training stages in MPT-Rec (as shown in Fig~\ref{fig:framework}). 

\begin{itemize}
\item \textbf{Multi-Task Pre-training Component.} 
Following the invariant learning~\cite{wang2022invariant}, which uses an environment classifier to identify the environment-variant and invariant information in an unsupervised way. 
In our pre-training component, we design a generative adversarial network with a task classifier to separate task-sharing and task-specific information, so as to ensure the high-quality transfer of task-sharing information in the prompt learning process of new tasks. 

\item \textbf{Multi-Task Prompt-tuning Component.}
In the prompt-tuning component, to avoid the negative transfer from the new task to existing tasks, MPT-Rec freezes the parameters in the pre-trained model and only updates the parameters of the new task with useful knowledge transferred from the pre-trained tasks by task-aware prompt mechanism.
\end{itemize}

In summary, we employ multi-task pre-training to acquire highly expressive task-sharing knowledge, ensuring MPT-Rec's predictive performance on existing tasks. 
With the high-quality transferred knowledge, we can further improve the learning performance on new tasks and gain an additional advantage in terms of training efficiency by freezing the parameters in pre-trained tasks.

\subsection{Multi-Task Pre-training}

The multi-task pre-training phase aims to enhance the performance of existing tasks while extracting transferable knowledge to facilitate the generalization of new tasks, which consists of two main operations: information disentangling and information fusion.

\subsubsection{Learning Disentangled Information}
To avoid the negative transfer problem among tasks, we use generative adversarial networks together with different experts to learn task-sharing and task-specific information respectively. 
The task-sharing information is expected to be well adapted to the new task, while the task-specific information needs to be further learned to provide useful information.
Follow the method in invariant learning~\cite{wang2022invariant}, which uses an environment classifier to identify the environment-variant and invariant information in an unsupervised way. 
In our pre-training component, we design a generative adversarial network with a task classifier to separate task-sharing and task-specific information. This ensures the high-quality transfer of task-sharing information in the prompt learning process. 
Specifically, the generative adversarial network consists of the task-sharing expert (as a generator) and the task classifier (as a discriminator). 
The goal of the generator is to confuse the task classifier by generating representations devoid of task-specific information.
Meanwhile, the discriminator aims to determine the task label associated with the task-shared representations. After training, the task-sharing expert will learn the shared information that can not be identified by the task classifier. 

Given the input raw features $\textbf{x}$, the input vector $\textbf{x}_{o} \in \mathbb{R}^{D}$ is learned from a sparse embedding network $\mathcal{I}$,  where $D$ is the dimension of the input vector. 
We assume that the function in the expert network is denoted as $\mathcal{F}$,
the task-shared representation $\textbf{x}_s \in \mathbb{R}^H$ and task-specific representations $\textbf{x}_k \in \mathbb{R}^H$ are denoted as  
$\textbf{x}_s=\mathcal{F}(\textbf{x}_o, \theta_s)$ and $\textbf{x}_k=\mathcal{F}(\textbf{x}_o, \theta_k)$, where $\theta_s$ and $\theta_k$ are the learning parameters in the expert networks, and $\mathcal{F}$ is a two-layer MLP (Multi-Layer Perceptron) with ReLU activation function.
Then we calculate the training loss of the generative adversarial network. The first loss is the prediction loss $Loss_s$ to ensure the predictive ability of task-sharing information, denoted by:
    \begin{equation}\label{eq2}
        \hat{y}_s = \mathcal{G}(\textbf{x}_s),  
    \end{equation}
    \begin{equation}\label{eq3}
        Loss_s = l_{rec}(\hat{y}_s, y_k),
    \end{equation}
where $\mathcal{G}$ represents a tower network composed of fully connected layers, 
$\hat{y}_s$ is the predicted value based on the task-sharing representations, and $y_k$ is the ground truth in the training dataset. For classification problems, $l_{rec}$ can be set to the widely used binary cross-entropy loss function.

The second loss in the generative adversarial network is the task label prediction loss $Loss_e$, defined as:
\begin{equation}\label{eq4}
    \hat{y}_e = \mathcal{K}(\textbf{x}_s, \phi),
\end{equation}
\begin{equation}\label{eq5}
    Loss_e = l_c(\hat{y}_e, y_e),
\end{equation}
where  $\mathcal{K}$ is the softmax learning function and $\phi$ is the learning parameters in the task classifier. $\hat{y}_e$ and $y_e$ represent the predicted value and ground truth of the task label respectively. As our datasets lack task labels, following~\cite{wang2022invariant}, we adopt the EM-based clustering algorithm to assign task labels and use the negative log-likelihood (NLL) loss function as $l_c$ to ensure the task label is correct. 

The final optimization loss for the generative adversarial network consists of the prediction loss $Loss_s$ and the $Loss_e$, defined as:
\begin{equation}\label{eq6}
    Loss_{gan} = \alpha Loss_e + (1 - \alpha) \sum_{k=1}^{N} Loss_s,
\end{equation}
where $\alpha$ is a weight coefficient used to balance the two loss functions, and $N$ is the number of tasks.

\subsubsection{Learning Fusion Information}
After separating the task-sharing and task-specific information through the generative adversarial network, we design a fusion network to combine them together for the final prediction of each task.  
Specifically, we assign a task embedding to each task to guide the fusion process.
The task-specific representation is combined with the task embedding to create a task-aware representation. 
Given a task $k$, the task-aware representation $\textbf{x}_e \in \mathbb{R}^{H}$ is denoted as:
    \begin{equation}\label{eq7}
        \textbf{x}_e = \textbf{x}_k \odot \textbf{E}_k,
    \end{equation}
where $\textbf{x}_k \in \mathbb{R}^{H}$ is the task-specific representation, $\textbf{E}_k \in \mathbb{R}^{H}$ represents the task embedding, and $\odot$ is the element-wise product operation. 

Subsequently, a gated network is employed to combine the task-sharing and task-aware representations in a weighted sum manner. For a task $k$, the weight coefficients $\beta_s$ and $\beta_e$ for the task-sharing and task-aware representations are automatically computed by the gated network $\mathcal{H}(\textbf{x}_o)$, where $\textbf{x}_{o} \in \mathbb{R}^{D} $ is the input vector from the embedding network, and $\mathcal{H}$ is one-layer MLP with softmax activation function.

The fusion representation $\textbf{x}_f \in \mathbb{R}^{H}$ can be formulated as:
    \begin{equation}\label{eq8}
        \textbf{x}_f = \beta_s \cdot \textbf{x}_s + \beta_e \cdot \textbf{x}_e.
    \end{equation}

Then the fused representation is forwarded to the upper tower network with a sigmoid function for task prediction. For task $k$, assuming that the prediction result of the fused representation is $\hat{y}_{f}$, the prediction loss $Loss_{f}$ for all tasks is computed as:
    \begin{equation}\label{eq9}
        Loss_f = \sum_{k=1}^N l_{rec} (\hat{y}_f, y_k).
    \end{equation}
Finally, the overall loss of the multi-task pre-training phase is obtained by summing the training loss for the generative adversarial networks and the prediction loss of the fused representations, defined as:
    \begin{equation}\label{eq10}
        \begin{split}
            Loss &= Loss_{gan} + Loss_{f}. 
        \end{split}
    \end{equation}
    
The multi-task pre-training component realizes the initial separation and subsequent fusion of knowledge through the tower network of each task, substantially mitigating the issue of negative transfer.
Simultaneously, task-sharing and task-specific representations, which are flexible in knowledge transfer, greatly enhance the generalization of new tasks.

\subsection{Multi-Task Prompt-tuning}
The multi-task prompt-tuning phase is designed to leverage existing knowledge, speeding up the training process for new tasks. This objective fits well in the recommendation situation where frequent requests are made for new task optimization due to the changes in business scenarios.
We propose a task-aware prompt-tuning method to extract useful information from other tasks. 
Task embeddings are used as prompts to combine the well-trained task-specific representations in the multi-task pre-training phase.

\subsubsection{Task-Specific Information Transfer}

Given an existing task $k$, the weight $\gamma_k$ to combine its specific information in the new task learning is computed as:
  \begin{equation}\label{eq11}
        \textbf{h}_{o} = \mathcal{P}(\textbf{x}_o),
    \end{equation}
    \begin{equation}\label{eq12}
        \gamma_k = \frac{e^{\textbf{h}_{o}\textbf{E}_k^\top}/T}{\sum_{k=1}^{N}e^{\textbf{h}_{o}\textbf{E}_k^\top}/T},
    \end{equation}
where $\mathcal{P}$ represents the projection function, which is a two-layer MLP. $\textbf{h}_{o} \in \mathbb{R}^{1 \times H} $ and $\textbf{E}_k \in \mathbb{R}^{1 \times H}$ are the vectors with same dimensions, and $T$ denotes a softmax temperature, that is used to scale the logits to prevent overconfidence.

By combining the useful information from existing tasks, the transferred representation $\textbf{x}_{t} \in \mathbb{R}^{H}$ for the new task  learning can be defined as: 
\begin{equation}\label{eq13}
        \textbf{x}_{t} = \sum_{k=1}^{N} \gamma_k \textbf{x}_k,
    \end{equation}
where $N$ is the number of tasks.

\subsubsection{Task-Aware Prompt Tuning }
We fuse the transferred task-specific representation $\textbf{x}_t$ with the new task embedding $\textbf{E}_n \in \mathbb{R}^{H}$ to obtain the task-aware representation $\textbf{x}_{new} \in \mathbb{R}^{H}$ of the new task, then combine it with the task-sharing representation $\textbf{x}_s $ for the new task prediction. 
The representations $\textbf{x}^{'}_{new} \in \mathbb{R}^{H}$  for the new task prediction is defined as as: 
\begin{equation}\label{eq14}
        \textbf{x}_{new} = \textbf{x}_t \odot \textbf{E}_n,
    \end{equation}
 \begin{equation}\label{eq15}
        \textbf{x}^{'}_{new} = \beta_s \cdot \textbf{x}_s + \beta_{new} \cdot \textbf{x}_{new}.
    \end{equation}
The optimization of the new task is formulated as:
   \begin{equation}\label{eq16}
        Loss_{new} = l_{rec} (\hat{y}_{new}, y_{new}),
    \end{equation}
where $\hat{y}_{new}$ is the prediction result from a sigmoid function with input vector $\textbf{x}^{'}_{new}$. 

\subsubsection{Efficiency Analysis}
Due to the comprehensive utilization of knowledge obtained during the multi-task pre-training phase. Our framework eliminates the necessity to re-train the sparse embedding network $\mathcal{I}$ of the high-dimensional input raw features, as well as the task-specific experts, which essentially reduces the training parameters and meanwhile improves the model efficiency. 
Only a small-scale projection network, tower network, and the new task embedding need to be learned in the prompt-tuning process. 
This efficient approach allows MPT-Rec to adapt to new tasks quickly. Moreover, we use task embeddings as prompts to guide weight learning, making it possible to transfer useful knowledge from the existing tasks to the new task. 
The multi-task prompt-tuning method ensures MPT-Rec's predictive capability on new tasks with minimal resource costs. These unique advantages make MPT-Rec an effective solution for addressing the challenge of new task generalization in recommender systems.

\begin{table}[thb]
\small
\caption{The statistics of datasets.}
\label{tab:Information about datasets}
\resizebox{\linewidth}{!}{
\begin{tabular}{cccc}
\toprule
Datasets & Train samples & Test samples & \#Feature \\
\midrule
Census-income & 199,523 & 99,762 & 40 \\
Ali-CCP & 42.3 million & 43 million & 18 \\
Byte-Rec & 15.7 million & 2 million & 8 \\
\bottomrule
\end{tabular}
}
\end{table}

\section{EXPERIMENTS}

\subsection{Experimental Settings}
\subsubsection{Datasets.} 
We conduct extensive experiments on three large-scale datasets, including two public datasets, Census-income and Ali-CCP, and a competition dataset ByteRec to validate the effectiveness of our proposed method in multi-task learning and its ability to generalize to new tasks. 
The statistics of the three datasets are summarized in Table~\ref{tab:Information about datasets}.

\begin{itemize}
\item Census-income~\cite{asuncion2007uci}. This dataset comprises census data extracted from the 1994 and 1995 population surveys.
Following prior work on multi-task learning, we designate \textbf{T1}: the prediction of whether an individual's income exceeds \$50,000, and \textbf{T2}: the prediction of their marital status. 

\item Ali-CCP~\cite{ma2018entire}. This dataset comprises 84 million samples collected from real-world traffic logs of the recommender system on Taobao. Following the settings in previous studies, we use \textbf{T1}: the prediction of the click-through rate (CTR) of items, and \textbf{T2}: the prediction of the conversion rate to evaluate the performance of MTL models. 

\item Byte-Rec~\cite{shen2021deepctr}. This dataset comes from a competition focused on short video recommendations.
It comprises tens of millions of interactions from tens of thousands of users. The dataset includes multi-modal short video content features and provides user interaction behavior data after desensitization. We evaluate the multi-task learning capability on two tasks. \textbf{T1}: predicting whether the user completed watching a video, and
\textbf{T2}: whether the user liked the video.
\end{itemize}

\begin{table*}[h]
\small
\caption{
    Experimental results of multi-task learning on three datasets. It is acknowledged by previous studies that a slight increase in AUC at 0.001 level is known to be a significant improvement in the MTL task. Besides, a t-test (with $p<0.05$, marked as "$\ast$") on the experiments shows the statistically significant improvements of our method over the best baseline (marked by underline). 
    The compared methods used to compute the gain of model performance are identified in italics.
}
\label{tab: Experimental results for multi-task training}
\begin{tabular}{c|cccc|cccc|cccc}
\toprule
\multirow{2}{*}{Approach} & \multicolumn{4}{c|}{Census-income} & \multicolumn{4}{c|}{Ali-CCP} & \multicolumn{4}{c}{Byte-Rec} \\
 & AUC/T1 & AUC/T2 & Gain/T1 & Gain/T2 & AUC/T1 & AUC/T2 & Gain/T1 & Gain/T2 & AUC/T1 & AUC/T2 & Gain/T1 & Gain/T2 \\
\midrule
Single Task & \emph{0.9443} & \emph{0.9834} & - & - & \emph{0.5730} & \emph{0.5903} & - & - & 0.7017 & 0.9340 & - & - \\
Shared Bottom & 0.9484 & 0.9897 & +0.0041 & +0.0063 & 0.5904 & 0.5860 & +0.0174 & -0.0043 & \emph{0.7292} & \emph{0.8442} & - & - \\
MMOE & 0.9494 & 0.9901 & +0.0051 & +0.0067 & 0.5881 & 0.5968 & +0.0151 & +0.0065 & 0.7289 & 0.8508 & -0.0003 &  +0.0066 \\
PLE & 0.9493 & 0.9911 & +0.0050 & +0.0077 & 0.5890 & 0.5998 & +0.0160 & +0.0095 & 0.7276 & 0.8605 & -0.0016 & +0.0163 \\
STEM & 0.9487 & 0.9906 & +0.0044 & +0.0072 & 0.5938 & 0.5983 & +0.0218 & +0.0080& 0.7385 & 0.9296 & +0.0093 & +0.0854 \\ 
Sparse Sharing & 0.9489 & 0.9914 & +0.0046 & +0.0080 & 0.5897& 0.6065& +0.0167& +0.0162& 0.7369 & 0.8923 & +0.0077 & +0.0481 \\
CSRec & {\ul 0.9502}& {\ul 0.9915} & {\ul +0.0059}& {\ul +0.0081} & {\ul 0.5913}& {\ul 0.6070}& {\ul +0.0183}& {\ul +0.0167}& {\ul 0.7412} & { \ul 0.8978} & {\ul +0.0120} & {\ul +0.0536} \\
MPT-Rec & \textbf{0.9517$^{\ast}$}& \textbf{0.9926$^{\ast}$} & \textbf{+0.0074}& \textbf{+0.0092} & \textbf{0.5990$^{\ast}$}& \textbf{0.6274$^{\ast}$} & \textbf{+0.0260} & \textbf{+0.0371} & \textbf{0.7485$^{\ast}$} & \textbf{0.9339$^{\ast}$} & \textbf{+0.0193} & \textbf{+0.0897} \\
\bottomrule
\end{tabular}
\end{table*}

\subsubsection{Baseline Methods.} 

We compare our proposed method MPT-Rec with several representative models in multi-task learning, including:

\begin{itemize}
\item Single Task. It optimizes each task individually without inter-task knowledge transfer.
\item Shared Bottom~\cite{ruder2017overview}. Multiple tasks share the same bottom network, followed by task-specific tower networks.
\item MMOE~\cite{ma2018modeling}. It replaces the bottom network in Shared Bottom with multiple expert networks to learn different aspects of knowledge. Additionally, a gated network is employed to learn the attention of different expert outputs.
\item PLE~\cite{tang2020progressive}. Following MMOE, PLE further divides the expert networks into task-sharing experts and task-specific experts.
\item STEM~\cite{su2023stem}. It is the latest work in the multi-task recommendation, in which simple separate embedding networks are used to learn the shared and task-specific embeddings. 
\item Sparse sharing~\cite{sun2020learning}. It utilizes pruning operations to learn a subnet separately for each task and then trains them in parallel. Knowledge is able to flow through overlapping parts between subnets.
\item CSRec~\cite{bai2022contrastive}. It is the SOTA baseline in multi-task learning. CSRec learns each task from separate subnets and constructs contrastive subnets to evaluate the contribution of parameters to a specific task, which alleviates the parameter conflict problem and enhances the optimization process.
\end{itemize}

Except for the Single Task model, the above methods cover different kinds of multi-task learning approaches in recommendation. Shared Bottom enables knowledge flow between tasks through the sharing bottom at the bottom. MMOE and PLE utilize expert information through gated networks. Sparse Sharing and CSRec design the models from the perspective of parameter efficiency.
The most similar models to our methods are STEM and PLE, both of them can learn task-sharing and task-specific knowledge at the embedding-level and expert-level respectively.  
Different from them, our proposed framework MPT-Rec is a two-stage multi-task prompt-tunning framework to deal with the negative transfer and high training costs problems in new task optimization in MTL recommendation.
MPT-Rec uses generative adversarial networks to separate the task-specific and task-sharing representations at the expert level, which enables our framework to achieve high training efficiency and good generalization ability in dealing with new tasks in recommender systems.

\subsubsection{Parameter Settings.} 
All methods are implemented in PyTorch with NVIDIA GeForce RTX 3090. 
For each baseline method, a grid search is applied to find the optimal settings. 
These include learning rate from $\{0.1, 0.01, 0.001, 0.0001, 0.00001\}$, hidden layer size of the expert network from $\{(256,192,128), (256,128), (128.64)\}$.
We report the result of each method with its optimal hyperparameter settings on the validation data. 
In our model, the dimension $D$ of input feature vectors after the embedding network is 127, 90, and 32 in Census-income, Ali-CCP, and Byte-Rec datasets respectively. The hidden layer size of the expert networks is set to (256, 128)in Census-income, and (128,64) in Ali-CCP and Byte-Rec datasets. The hidden layer sizes of the projection network are (64,128), (32,64), (32,64) in three datasets. The feature embedding dimension $H$ is 128, 64, 64, and the learning rate is 1e-3, 1e-4, 1e-4 in three datasets respectively. 
The balance weight $\alpha$ for the two loss terms in the generative adversarial network training to 0.1.
The code will be publicly available after the review process.

\subsection{Results of Multi-Task Learning} 
The results on three datasets are shown in Table~\ref{tab: Experimental results for multi-task training}. 
All tasks on three datasets are binary classification tasks, we use AUC as the evaluation metric. We have the following observations:

(1) Compared with the Single Task model, all the multi-task models achieve a certain performance gain except for the T2 in Byte-Rec dataset, which indicates that the joint training of multiple tasks can promote the transfer of related knowledge among tasks, which can promote the model performance of each task. 

(2) For the multi-task learning models, the Shared Bottom model performs the worst, because the task-irrelevant knowledge from other tasks may also be learned in the shared bottom network. MMOE and PLE perform better than the Shared Bottom model, they use the gated network to filter the useful information with different experts. 

(3) The performances of the STEM model on different datasets are not consistent: it performs relatively better than other baselines in task T2 on the Byte-Rec dataset but loses advantages compared with PLE in the Census-income dataset. It indicates that distinguishing the task-specific and sharing information by simple embedding networks can not well address the negative transfer problem. 

(4) Sparse Sharing and CSRec learn different subnets for each task, and only the overlapping subnets allow knowledge transfer. Thus, they achieve better experimental performance with fewer parameters. In addition, CSRec uses contrastive learning to solve the problem of parameter conflict and further improves the model performance.

(5) Our proposed method MPT-Rec achieves the best performance on all datasets. Different from PLE, which uses task-sharing and task-specific experts to distinguish the task-sharing and task-specific information, MPT-Rec uses a generative adversarial network to make more explicit restrictions in the learning process, making it more capable of avoiding the transfer of task-irrelevant information.

\begin{table*}[]
\small
\caption{Experimental results of new task generalization on Census-income and Ali-CCP datasets. \#Params denotes the number of trainable parameters and \#FLOPs denotes the number of floating-point operations per batch. We report the performance of MTL methods in the prompt-tuning scheme, e.g., Shared$^{*}$, MMOE$^{*}$, PLE$^{*}$. Besides, comparing training them in the full-training scheme (e.g., Shared, MMOE, PLE), we report the percentage of \#Params and \#FLOPs in parentheses. The fine-tuning scheme inevitably suffers from performance degradation because only a fraction of the parameters are learned. Compared with MPT-Rec in a full-training scheme, our prompt-tuning model remains 99.7\% and 95.5\% model performance on two datasets.}
\label{tab:Experimental Results on generalization to new tasks}
\begin{tabular}{>{\centering\arraybackslash}p{\dimexpr 0.15\linewidth-2\tabcolsep}
                >{\centering\arraybackslash}p{\dimexpr 0.15\linewidth-2\tabcolsep}
                >{\centering\arraybackslash}p{\dimexpr 0.15\linewidth-2\tabcolsep}
                >{\centering\arraybackslash}p{\dimexpr 0.15\linewidth-2\tabcolsep}
                >{\centering\arraybackslash}p{\dimexpr 0.15\linewidth-2\tabcolsep}
                >{\centering\arraybackslash}p{\dimexpr 0.15\linewidth-2\tabcolsep}}
\toprule
Approach & MAML & Shared$^{*}$ & MMOE$^{*}$ & PLE$^{*}$ & MPT-Rec \\
\midrule
\multicolumn{6}{c}{Census-income} \\ \midrule
AUC/T3 & 0.8563 & 0.8133 & 0.8497  & 0.8617  &  \textbf{0.8626} (99.7\%) \\
\#Params & 26k & 21k (21.2\%) & 21k (9.4\%) & 118k (26.0\%)  & \textbf{27k} (9.3\%) \\
\#FLOPs & 19M & 3M (10.8\%) & 3M (4.7\%) & 30M (26.2\%) & \textbf{7M} (8.7\%)  \\
\midrule
\multicolumn{6}{c}{Ali-CCP} \\ \midrule
AUC/T3 & 0.7155 & 0.6279 & 0.6862 & 0.6997 & \textbf{0.6968} (95.5\%) \\
\#Params & 1748k & 6k (0.12\%) & 7k (0.13\%) & 33K (0.62\%)  & \textbf{8k} (0.15\%) \\
\#FLOPs & 43M & 6M (11.2\%) & 7M (5.1\%) & 66M (26.4\%)  &  \textbf{17M} (9.2\%)  \\
\bottomrule
\end{tabular}
\end{table*}

\subsection{Results of New Task Generalization}
To evaluate the generalization ability of MTL methods, we construct a new task T3 by excluding the predictive feature. Due to the space limitation, we present the experimental results using the feature "education" in the Census-income dataset and the "business scenario" in the Ali-CCP dataset as T3. 

\subsubsection{Negative Transfer on Full-training Scheme}
We first verify the negative impacts on existing tasks by incorporating the new task into MTL models.
Existing MTL models, e.g., MMOE, PLE, adopt \emph{full-training scheme} to learn with the new task: all three tasks are trained simultaneously, and all parameters are updated in the entire model. 
Considering that the new task's impacts on existing tasks may be influenced by their correlations, we compute the Pearson's correlation coefficient to measure the similarity of distributions between them and test on two features (i.e., "Sex" and "Education") with different correlation coefficient in the Census-income dataset. 
As shown in Table~\ref{tab negative impacts}, 
we can see that 
for the new task T3 "Education" (\emph{correlation coefficient T3-T1: 0.186, T3-T2: 0.142}) with a relatively small gap in average correlation coefficient to the existing tasks (i.e., \emph{the correlation coefficient of T1-T2 is 0.178}), only MMOE will gain benefits from the joint training on existing tasks, most of MTL models suffer the negative impacts on existing tasks. 
For the new task T3 "Sex" (\emph{correlation coefficient T3-T1: 0.158, T3-T2: 0.063}), which has a lower correlation coefficient, the incorporation of new tasks will negatively impact existing tasks in all models. 
We can see that for the new tasks with different data distribution, it is necessary to design a mechanism to separate the task-sharing and task-specific information, so as to alleviate the negative transfer of irrelevant information from other tasks and boost the model's performance.
As the generalization ability of each MTL model may be different,
for fair comparisons, we report the new task (T3: "Education") with a similar correlation coefficient as the previous two tasks T1 and T2 in our following experiments.

\begin{table}[htb]
\caption{The positive or negative impacts on the existing task T1 and T2 after making full training with T3 in multi-task learning methods. "-" means the negative impact and "+" means the positive impact. Avg.Coef is the average correlation coefficient between new task T3 and tasks T1, T2. }
\label{tab negative impacts}
\begin{tabular}{l c|cccc}
\toprule
 \multicolumn{6}{c}{Model} \\
 \multicolumn{2}{c}{New task}  & Shared & MMOE & PLE & STEM \\
\midrule
T3: Education & T1 & \fontsize{12}{12}\selectfont + & \fontsize{12}{12}\selectfont + & \fontsize{12}{12}\selectfont - & \fontsize{12}{12}\selectfont -  \\
Avg.Coef:0.16  & T2 & \fontsize{12}{12}\selectfont - & \fontsize{12}{12}\selectfont + & \fontsize{12}{12}\selectfont + & \fontsize{12}{12}\selectfont - \\ \hline
T3: Sex & T1 & \fontsize{12}{12}\selectfont - & \fontsize{12}{12}\selectfont - & \fontsize{12}{12}\selectfont - & \fontsize{12}{12}\selectfont -  \\
Avg.Coef:0.11 & T2 & \fontsize{12}{12}\selectfont - & \fontsize{12}{12}\selectfont - & \fontsize{12}{12}\selectfont - & \fontsize{12}{12}\selectfont -  \\
 \bottomrule
\end{tabular}
\end{table}

\subsubsection{Efficiency Improvements in Fine-tuning Scheme}
\begin{figure}
\centering
\setcounter{subfigure}{0}
\subfigure[Shared*]{\label{fig:sharedbottom}
\includegraphics[height=0.37\linewidth]{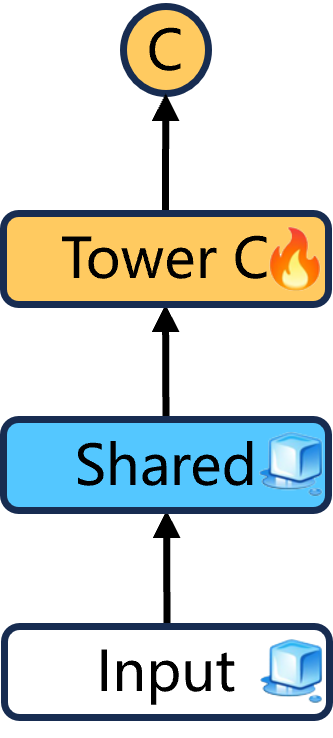}}
\hspace{0.01\linewidth}
\subfigure[MMOE*]{\label{fig:mmoe}
\includegraphics[height=0.36\linewidth]{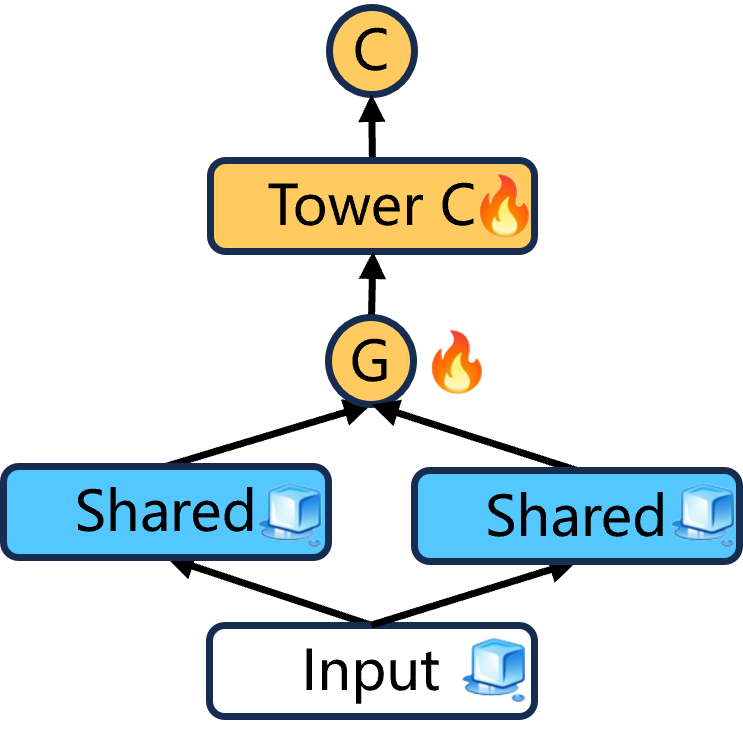}}
\hspace{0.01\linewidth}
\subfigure[PLE*]{\label{fig:ple}
\includegraphics[height=0.36\linewidth]{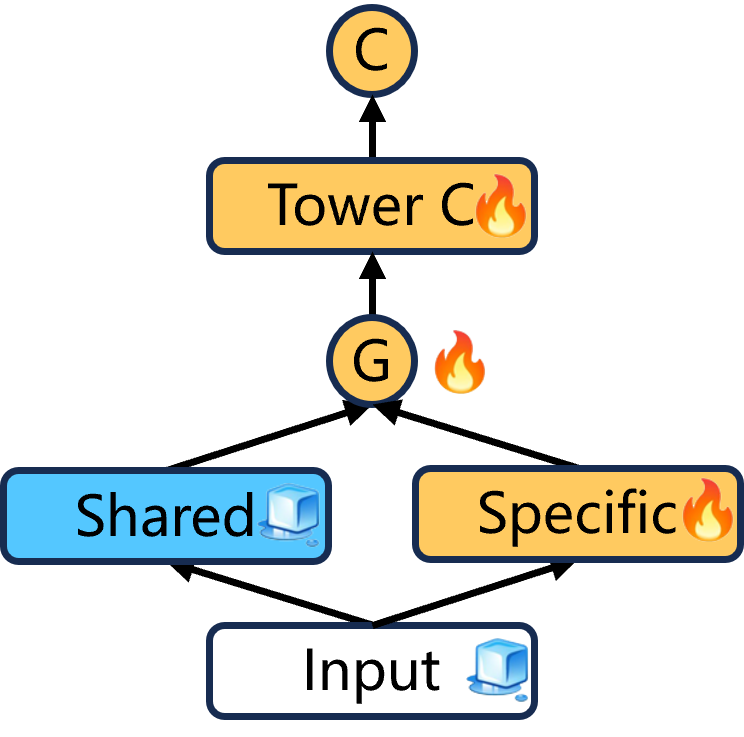}}
\caption{Fine-tuning operations on Shared Bottom, MMOE and PLE methods. "fire" means the parameters need to be trained, and "ice" indicates the freeze status.}
\label{fine-turning methods}
\end{figure}

We further evaluate the generalization ability of our model in dealing with the new task in the fine-tuning scheme. 
To make a fair comparison, we freeze the feature embedding networks in baseline methods and optimize their upper tower networks for the new task. 
The fine-tuning operations for the typical MTL methods are shown in Fig.~\ref{fine-turning methods}. The compared methods for new task learning in MTL recommendation are as follows:
\begin{itemize}
    \item Shared Bottom*. It freezes the shared network at the bottom.
    \item MMOE*. It freezes all expert networks, while the gated network that learns attention can still be trained.
    \item PLE*. It freezes all shared expert networks, while task-specific experts and gating networks can still be trained.
    \item MAML~\cite{finn2017model}. It is a typical model-agnostic method by mete-learning to address the new task adoption problem.  
\end{itemize}

The experimental results of different methods are shown in Table~\ref{tab:Experimental Results on generalization to new tasks}. We have the following observations:

(1) The model-agnostic method MAML performs the worst in experiments because the information can not be fully shared by the mete-learning mechanism. PLE* performs better than Shared Bottom* and MMOE*. Because Shared Bottom and MMOE do not model the general information explicitly, the transferred knowledge is not well adapted to the new task. 

(2) Our MPT-Rec further improves performance over PLE*, because the sharing information learned in PLE is mixed with more noise that is useless for the new task, while MPT-Rec learns high-quality task-sharing information by using generative adversarial networks. Compared to the SOTA method PLE*, it shows a particular advantage in terms of parameters and FLOPs.


(3) Compared with the full training scheme, the fine-tuning scheme greatly improves the training efficiency although at the cost of a certain performance degradation. 
In particular, our proposed MPT-Rec achieves more than a 90\% reduction in the number of parameters and FLOPs on both datasets. As for the model performance, compared to MPT-Rec in a full-training scheme, our prompt-tuning model remains 99.7\% in the Census-income dataset and 95.5\% in the AliCCP dataset respectively.

(4) The reduction of parameters in the Ali-CCP dataset is much smaller than that of the Census-Income dataset, which indicates that the larger the dataset is, the more obvious the advantage of our proposed fine-tuning scheme in multi-task learning. This is because as the increasing of dimension in the input feature, the parameters in the feature embedding network of the corresponding model will expand rapidly, showing the necessity to learn the new task in a fine-tuning scheme.

In summary, our proposed multi-task prompt-tuning approach provides a better trade-off between model performance and resource costs. It is able to make full use of the knowledge obtained in the multi-task pre-training phase, which can greatly accelerate the learning of new tasks and avoid the negative impacts on existing tasks.
It can be well fit for large-scale real-world recommender systems, especially in situations where frequent requests are made for new task optimization due to the changes in business scenarios.


\begin{figure}[h]
\centering
\setcounter{subfigure}{0}
\subfigure[\textbf{Census-income}]{\label{fig:Share and Specific(1)}
\includegraphics[width=0.42\linewidth]{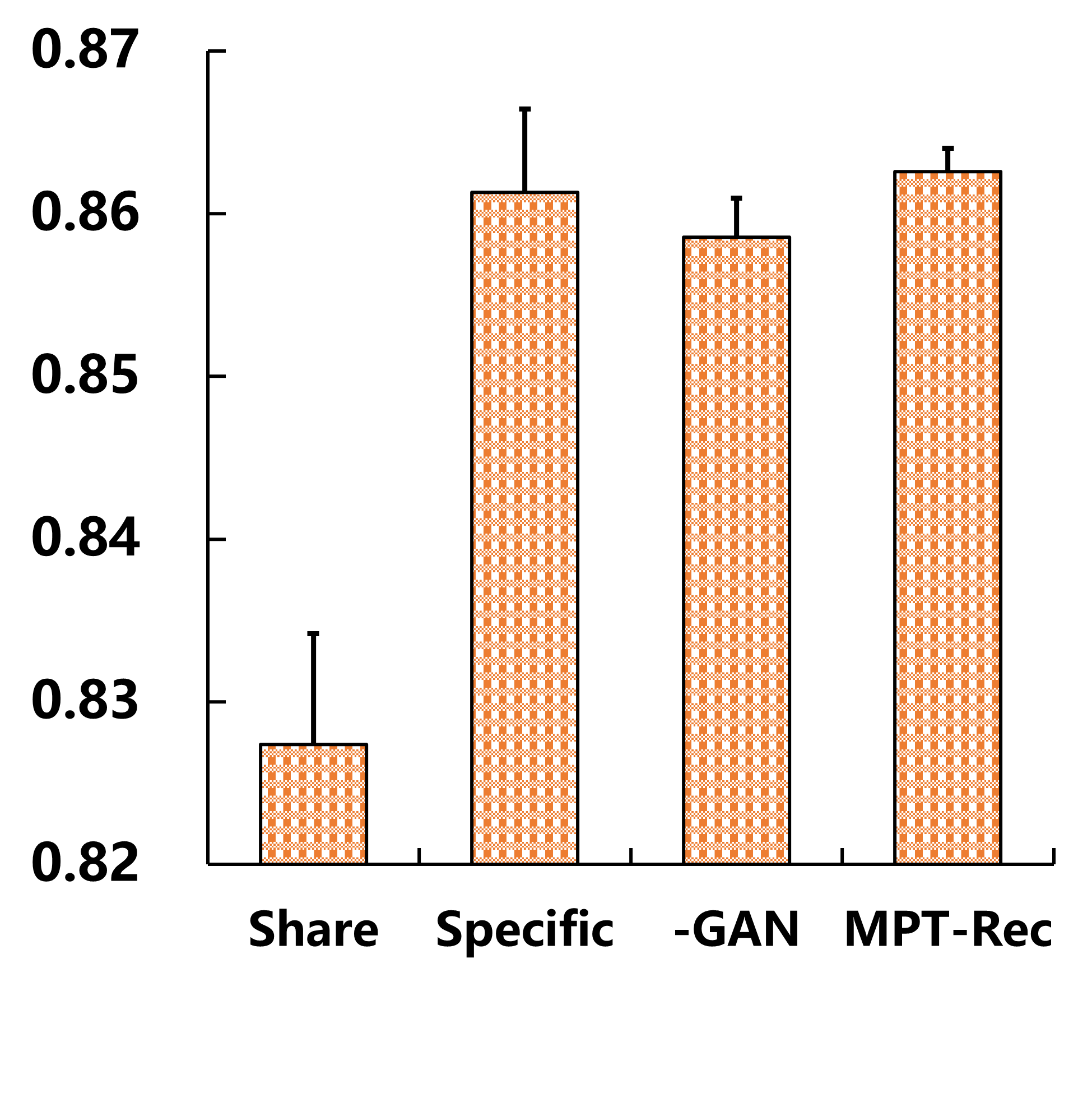}}
\hspace{0.01\linewidth}
\subfigure[\textbf{Ali-CCP}]{\label{fig:Share and Specific(2)}
\includegraphics[width=0.42\linewidth]{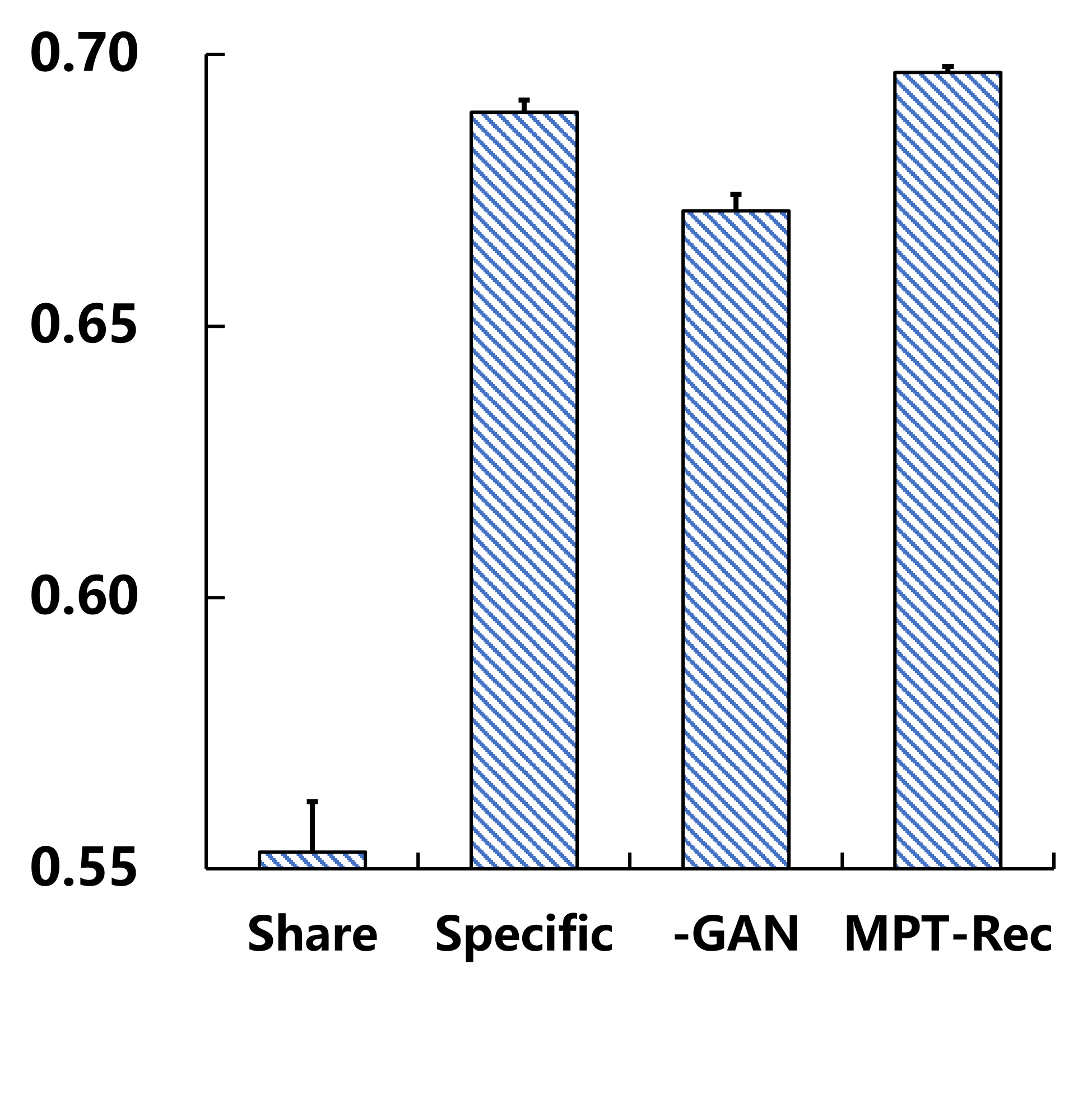}}
\caption{ Performance comparison of variant models on new task prediction on Census-income and Ali-CCP datasets. "Share" and "Specific" denote only task-shared and task-specific information is used. "-GAN" refers that does not disentangle the task-specific and task-sharing information by removing the GAN part.}
\label{fig:Share and Specific}
\end{figure}

\begin{figure}[]
\centering
\setcounter{subfigure}{0}
\subfigure[\textbf{Census-income}]{\label{fig:FW and TES(1)}
\includegraphics[width=0.42\linewidth]{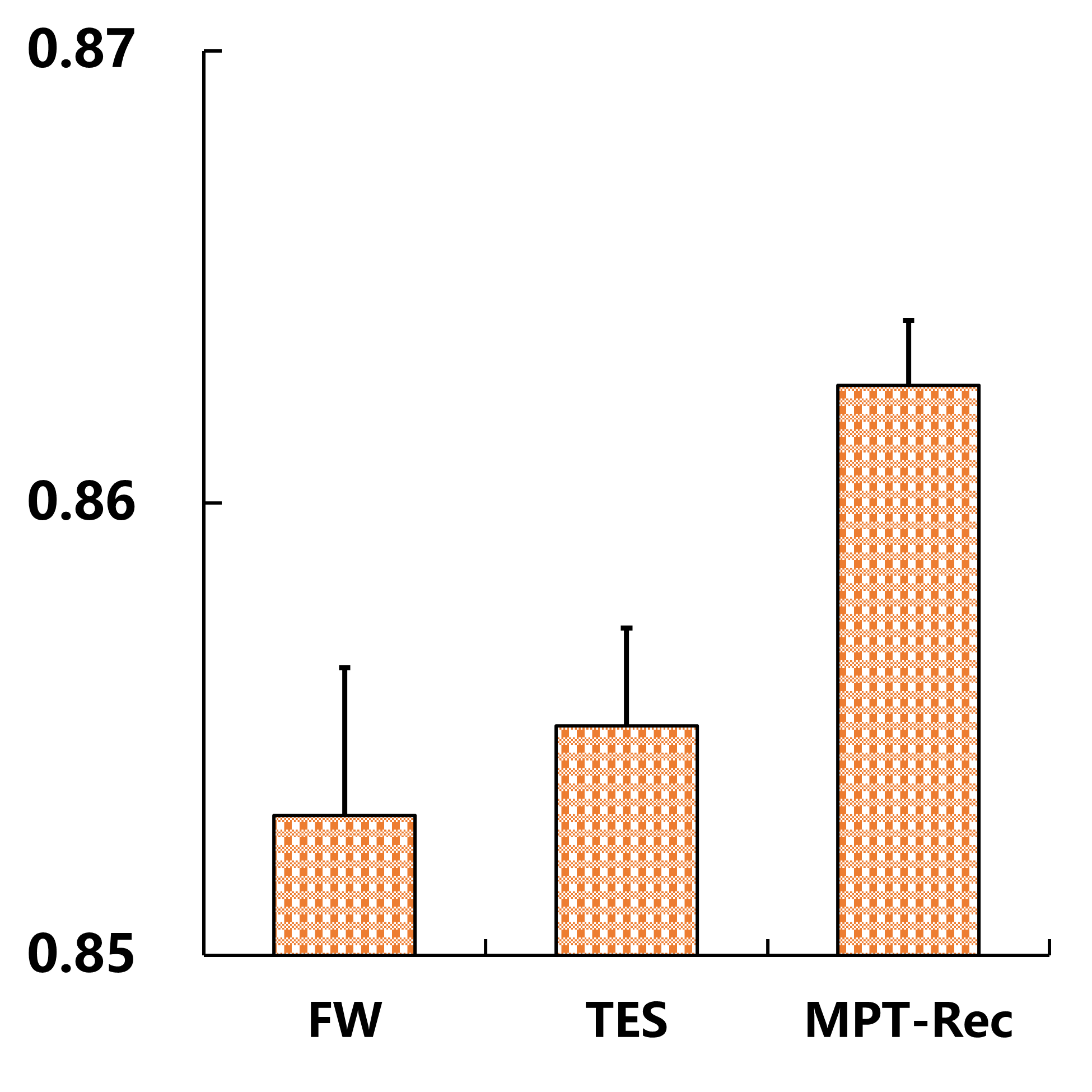}}
\hspace{0.01\linewidth}
\subfigure[\textbf{Ali-CCP}]{\label{fig:FW and TES(2}
\includegraphics[width=0.42\linewidth]{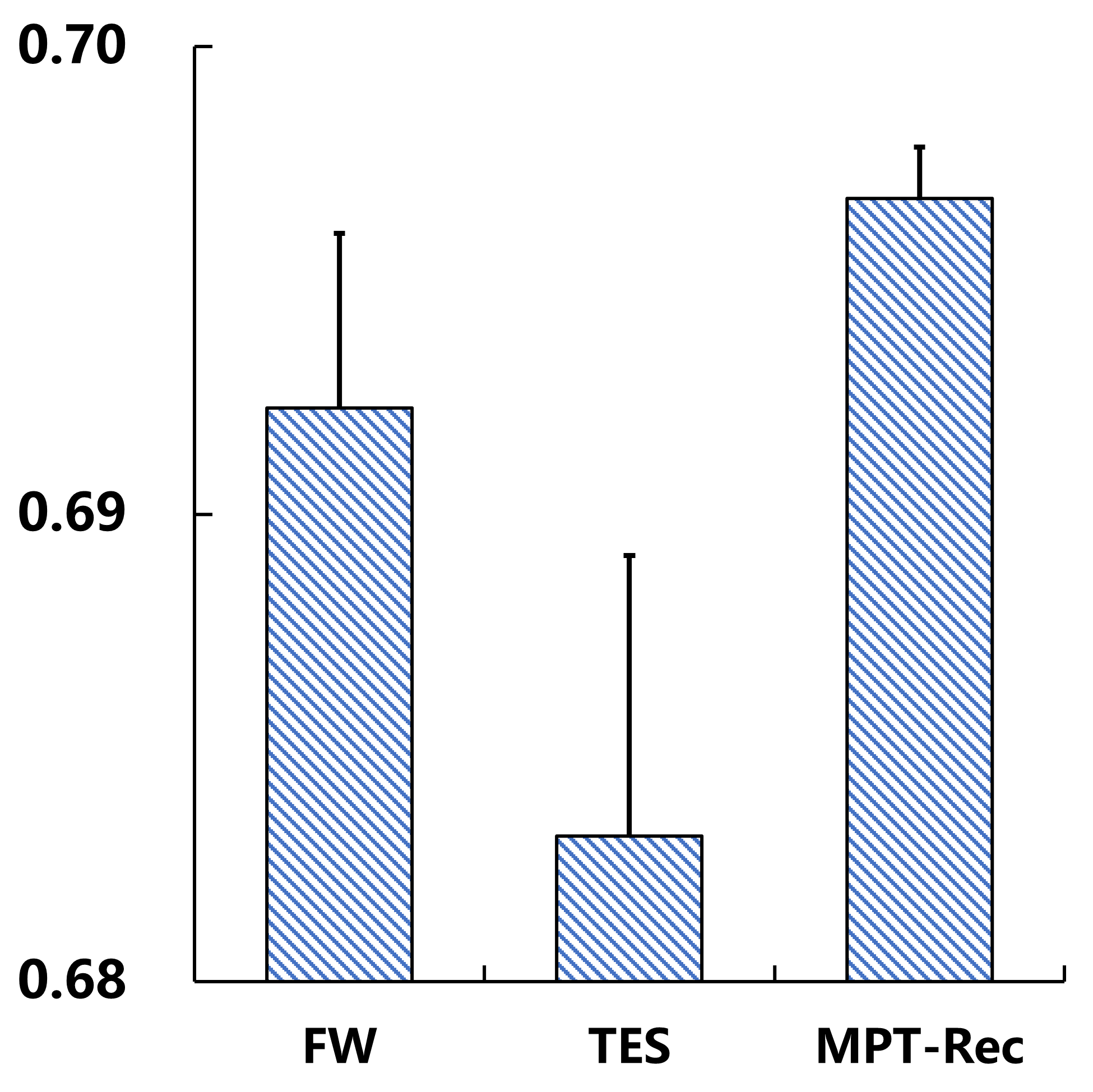}}
\caption{Performance comparison of different methods to combine the specific information from other tasks.
"FW" uses fixed weights to fuse the task-specific knowledge in the pre-training phase.
"TES" calculates the fusion weights according to task embedding similarity.}
\label{fig:FW and TES}
\end{figure}

\begin{figure*}[]
\centering
\setcounter{subfigure}{0}
\subfigure[\textbf{Census-income}]{\label{fig:Census-income-visualization}
\includegraphics[width=0.88\linewidth]{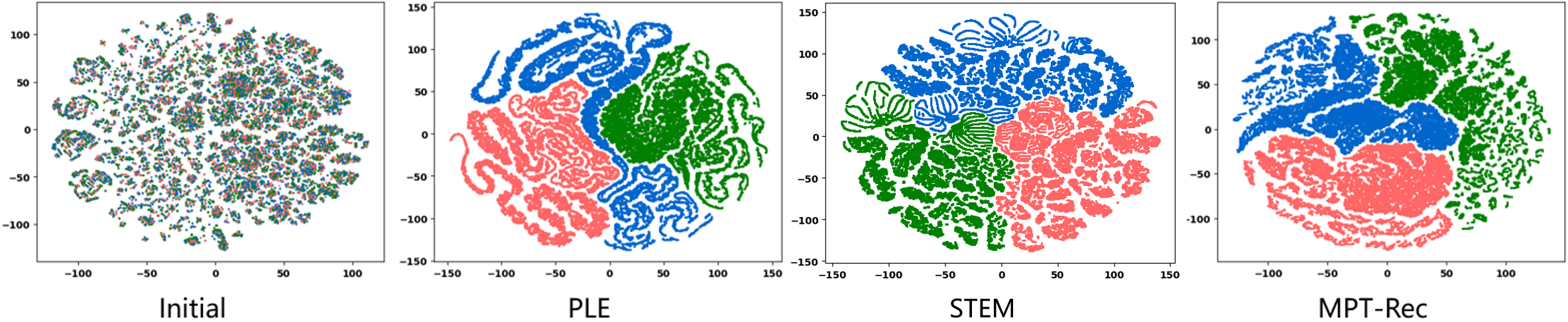}}
\subfigure[\textbf{Ali-CCP}]{\label{fig:Ali-CCP-visualization}
\includegraphics[width=0.88\linewidth]{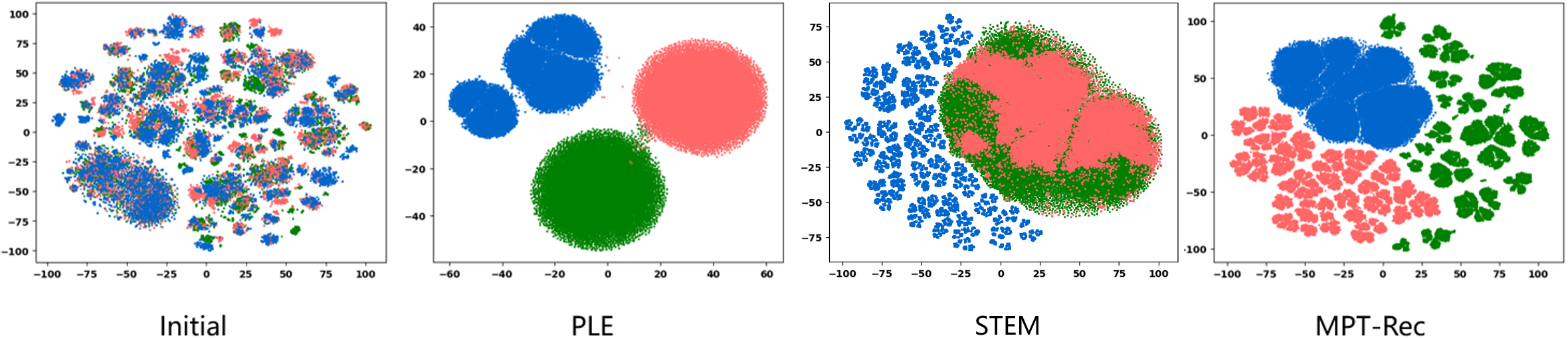}}
\caption{The Visualization of disentangled information. The green color marks the task-sharing information, the different task-specific information is marked by the blue and pink colors respectively.}
\label{fig:visualization}
\end{figure*}

\subsection{Experimental Analysis}
In this section, we conduct ablation studies to show the effectiveness of the multi-task prompt-tuning component. Then we further make visualization of the task-sharing and task-specific information for better explainability.

\subsubsection{The Effectiveness of the Disentangled Pre-training} 
To verify the effectiveness of the task-aware generative adversarial network in learning the task-sharing and task-specific information in the pre-training component of MPT-Rec, we conduct the ablation experiment with two variants, i.e., MPT-Rec (Share) and MPT-Rec (Specific), which only utilize the task-sharing information and the task-specific information respectively. 
We also remove the generative adversarial component, termed MPT-Rec (-GAN), in which both the task-shared and task-specific information are mixed together.
We test the AUC scores of the three variant models on the Census-income and Ali-CCP datasets. 
The results are shown in Fig~\ref{fig:Share and Specific}. We can see that the performance of both MPT-Rec (Share) and MPT-Rec (Specific) decrease compared with MPT-Rec, indicating that both information is useful in the new task prediction. The decrement of model performance is much larger in the MPT-Rec (Share) variant, which may be caused by the task conflict problem, where only a little information is shared among them. In this case, the task-specific information is more beneficial to the new task learning and it is necessary to extract the useful information from it. 
Without the separation process by GAN, the variant MPT-Rec(-GAN) performs worse than the specific model with only part of the information, showing the effectiveness of using the generative adverisal networks in our framework.

\subsubsection{The Effectiveness of the Prompt-tuning Mechanism} 
In MPT-Rec, we use a task-aware prompt mechanism to extract the information from other tasks. To verify the effectiveness of our prompt operation, we compare MPT-Rec with two variants, i.e. MPT-Rec (FW) and MPT-Rec (TES). 
MPT-Rec (FW) uses fixed weights (e.g., equal weights) as hyperparameters to fuse the task-specific information and task-sharing information obtained in the multi-task pre-training phase.
MPT-Rec (TES) calculates the fusion weights according to Task Embedding Similarity, which is calculated by the dot product of the embeddings between the new task and the existing task.
The difference between MPT-Rec and MPT-Rec (TES) is that MPT-Rec can extract the task-specific knowledge at an instance level, while in MPT-Rec (TES), the knowledge is extracted at a task level (i.e., the weights of all instances in a task are the same).
The experimental results are shown in Fig~\ref{fig:FW and TES}. Our task-aware prompt-tuning scheme achieves the best performance since the other two reconstruction operations are conducted at the task level.
Our prompt learning operation is more flexible and capable of learning fine-grained fusion weights, i.e., it learns fusion weights for each input instance.

\subsubsection{The Visualization of Disentangled Information}
We further visualize the task-sharing and task-specific representations learned in MPT-Rec. 
All the representations are randomly initialed and trained in the multi-task learning models. We utilize t-SNE to map the high-dimensional representation vectors to a two-dimensional space and report our visualization results in Fig~\ref{fig:visualization}. The green color marks the task-sharing information; the different task-specific information is marked by the blue and pink colors respectively. 
We can see that after training, our MPT-Rec shows clear boundaries between the three types of information. Compared with PLE, the boundary between two task-specific representations is more evident in the Census-income dataset. For the Ali-CCP dataset, STEM loses the advantage of distinguishing the task-sharing information, which is confused with the task-specific information. Both PLE and our method distinguish the three types of information well. For the task-sharing representation, the distance is more closer to the task-specific information in our model, which helps to make a fusion of the information and further improve the model performance.

\section{conclusion}
In this paper, we propose an effective multi-task learning framework and also investigate the generalization problem of multi-task learning in dealing with new tasks. We point out the negative transfer and high resource cost problems in the new task learning in recommender systems. 
We propose a novel two-stage (i.e., pre-training and prompt-tuning stages) multi-task learning framework, MPT-Rec. It separates task-sharing and task-specific information in the pre-training stage and allows it to be fully utilized in the prompt-tuning stage. 
Our proposed method MPT-Rec effectively solves the negative transfer in multi-task optimization and high-cost problems in new task learning, which provides a solution to make trade-offs between model performance and resource overhead in large-scale real-world recommender systems.

\bibliographystyle{ACM-Reference-Format}
\bibliography{sample-base}

\end{document}